# Natural Language Processing and Query Expansion[1]
## Issues, state-of-the-art and perspectives

B. Selvaretnam & M. Belkhatir
Faculty of IT, Monash University and University of Lyon & CNRS, France

**Abstract**

The availability of an abundance of knowledge sources has spurred a large amount of effort in the development and enhancement of Information Retrieval techniques. Users' information needs are expressed in natural language and successful retrieval is very much dependent on the effective communication of the intended purpose. Natural language queries consist of multiple linguistic features which serve to represent the intended search goal. Linguistic characteristics that cause semantic ambiguity and misinterpretation of queries as well as additional factors such as the lack of familiarity with the search environment affect the users' ability to accurately represent their information needs, coined by the concept intention gap. The latter directly affects the relevance of the returned search results which may not be to the users' satisfaction and therefore is a major issue impacting the effectiveness of information retrieval systems. Central to our discussion is the identification of the significant constituents that characterize the query intent and their enrichment through the addition of meaningful terms, phrases or even latent representations, either manually or automatically to capture their intended meaning. Specifically, we discuss techniques to achieve the enrichment and in particular those utilizing the information gathered from statistical processing of term dependencies within a document corpus or from external knowledge sources such as ontologies. We lay down the anatomy of a generic linguistic based query expansion framework and propose its module-based decomposition, covering topical issues from query processing, information retrieval, computational linguistics and ontology engineering. For each of the modules we review state-of-the-art solutions in the literature categorized and analyzed under the light of the techniques used.

## 1  Introduction

Over the years, the transition into the information age has brought about a tremendous growth in the information space, both online and offline. The birth of knowledge-based economies has further spurred the escalation of a vast array of knowledge sources available for manipulation. The very nature of the rapidly changing information space and the diverse search goals of users, present an inherent challenge to the task of information retrieval. Information Retrieval is the act of finding materials, be it text, audio or video, from within large collections to satisfy an information need. An information need is expressed in natural language and successful retrieval is very much dependent on the effective communication of the intended purpose.

Current commercial search engines are based on keyword indexing systems and Boolean logic queries. Keyword lists describe contents of documents or information objects but do not say anything about relationships (either semantic or statistical) between keywords. The most significant problem with keyword-based information retrieval is that it performs string matching and thus it does not take into account the correlations between words or phrases. In lieu of this, much research

---

[1] This version is a preprint sent to the Journal of Intelligent Information Systems. The full reference is:
*B. Selvaretnam, M. Belkhatir. Natural language technology and query expansion: issues, state-of-the-art and perspectives. J. Intell. Inf. Syst. 38(3): 709-740 (2012)*

efforts have emerged in the development of concept-based information retrieval. With concept based information retrieval, the search for information is based on the meaning of words or phrases rather than merely the presence of keywords within the index. Typically, a query that specifies the intent of a user's search goal is provided to a retrieval system (e.g. web search engines, specialized search systems etc) with the hopes of obtaining the most relevant set of results for an information need. Given the large information space in online retrieval systems as well as the fact that a lot of valuable information is lost in translation when users try to represent their actual information need in the form of a query (also coined as the **intention gap**), a search query quite often results in a long list of results being returned and this requires a tedious evaluation of the huge amount of potential resources obtained.

To illustrate, search results obtained by running a query on the Google search engine are discussed below.In a first scenario, Edwin's class has been assigned the task of preparing a poster on physical resources that are useful for people who have disabilities or the lack there of. Physical resources here would entail things like crutches, wheelchairs, ramps, handrails, designated parking spaces etc. With this in mind, Edwin performs a search with the query "physical resources for disabled people". To his dismay, none of the top 25 images satisfied his need (cf. figure 1). He then proceeds to enter another i.e. query "disabled people resources". However, this time he was able to obtain two images that were acceptable for his need (Figure 2). On the other hand, Lisa, who had been given a similar task, was successful in obtaining many more relevant images (Figure 3) within the top 25 images and beyond. Lisa had submitted the query "handicapped people facilities". In this instance, although both Edwin and Lisa had the same goal in mind, they had both expressed their information need by composing queries with differing structures and word combinations, thus resulting in different levels of accuracy in their results.

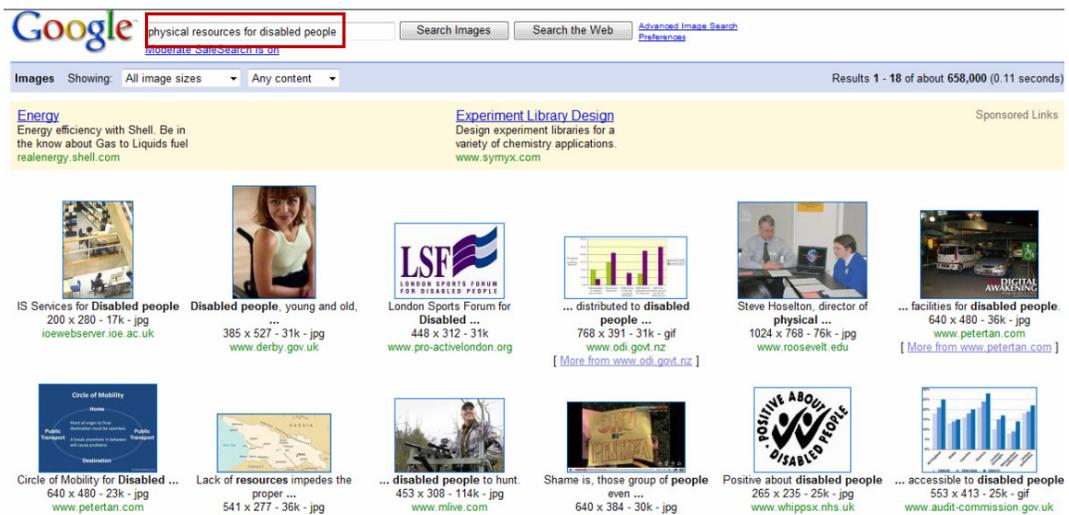

**Figure 1. Search Results for "physical resources for disabled people"**

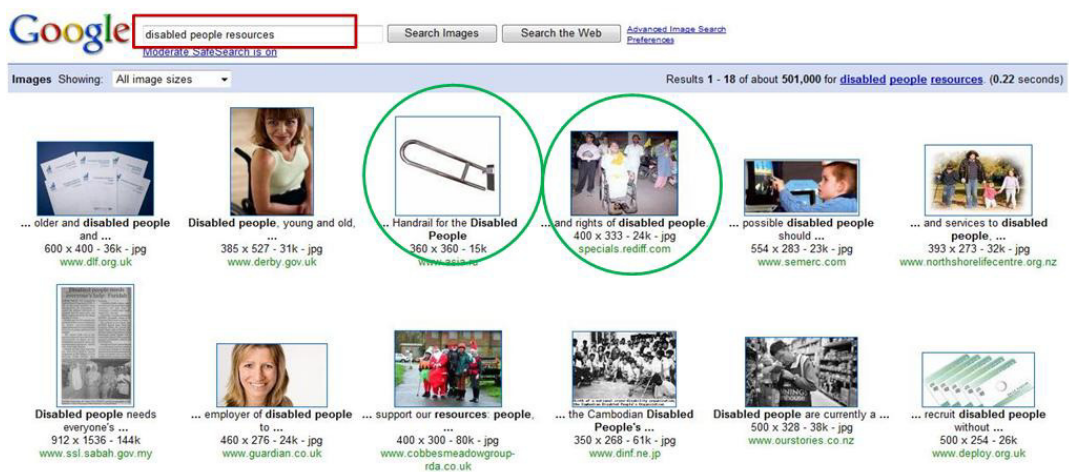

Figure 2. Search Results for "disabled people resources"

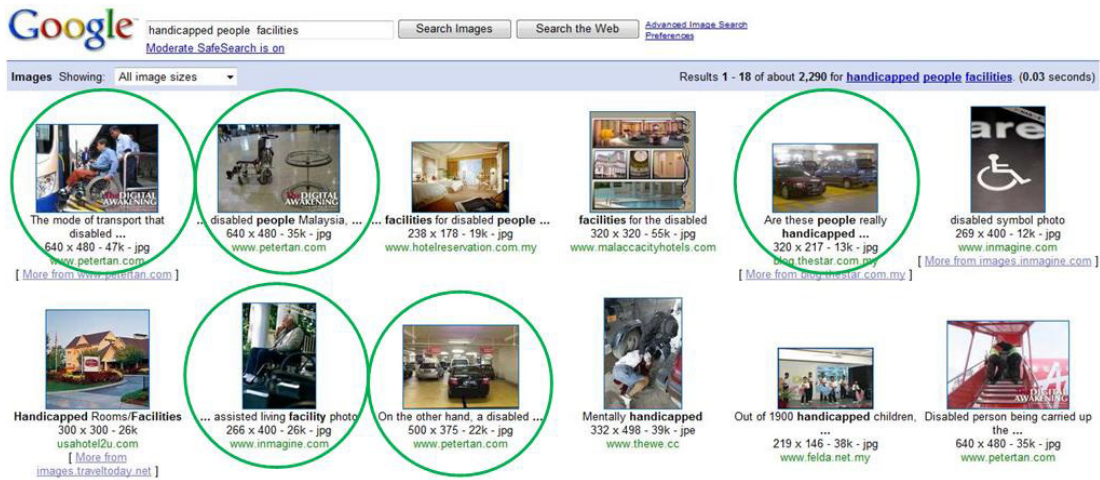

Figure 3. Search Results for "handicapped people facilities"

In the second scenario, we assume a user, Aleya, is looking to buy her first vehicle. Not knowing for sure what type of transport to buy (e.g. car, van, motorcycle), she decides to conduct a survey of available vehicles on sale before making her decision. She proceeds to submit a query that reads "vehicles for sale". Figure 4 (Query A) displays the results of her search. The results returned are indeed relevant. However, the term "vehicle" in the English language can be represented as "automobile". Simply by replacing the term "vehicle" with its synonym "automobile", another set of relevant results can be obtained as shown in Figure 4 (Query B). When the top 5 results of both search runs are closely observed, we note that there is only one site that matches in both results set (e.g. www.autotrader.com). None of the other sites appear even in the next 50 results of her search. This would mean that Aleya is missing out on some relevant sites which may have been more pertinent to her search due to her choice of search terms.

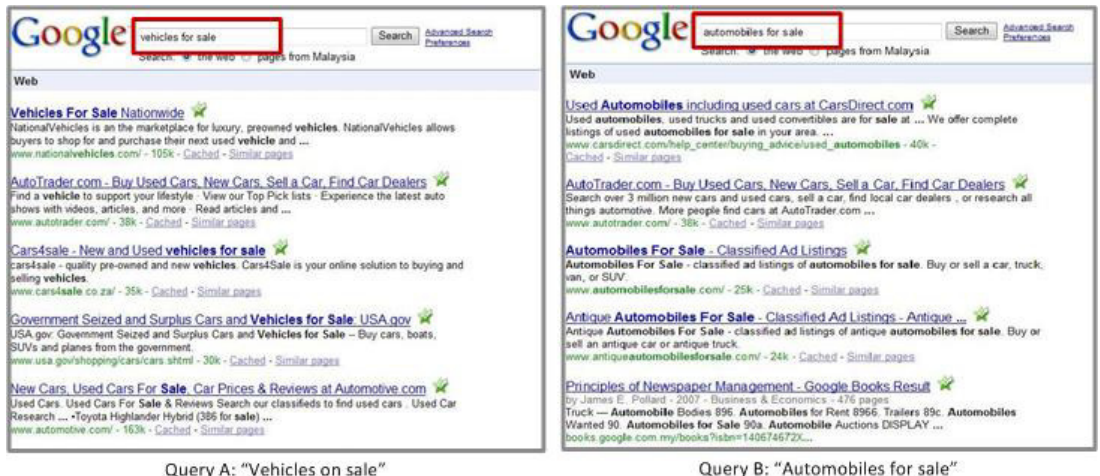

Figure 4 Search Results for "Vehicles for sale" & "Automobiles for sale"

The scenarios discussed above illustrate the *query-document vocabulary mismatch problem* that has plagued the Information Retrieval domain. Often times, it is observed that the accuracy of query formulation depends on a user's familiarity with search tasks and of the subject matter. Users are plagued with the inability to accurately represent their information needs due to a number of factors which are, among other things, lack of familiarity with domain specific vocabulary, lack of clarity in expressing their intent, lack of familiarity with the search environment and minimal awareness of the logic applied on to their query when a search request is made (Salton & Buckley, 1990; Spink et al., 1998; White & Morris, 2007). Consequently, the returned search results are very often not to the users' satisfaction which then leads to the reformulation of their initial query. This issue may be addressed through query reformulation, the process of modifying a seed query provided by the user. The process of query expansion (QE) would allow user given queries to be enhanced to capture its intended meaning and as a result improve precision (i.e. the fraction of the documents retrieved that are relevant to the user's information need) and recall (i.e. the fraction of the documents that are relevant to the query that are successfully retrieved). There is a need to distinguish between two terminologies, query reformulation and query expansion, that are utilized to represent the process of modifying queries. Query reformulation involves either the restructuring of the original query or by adding new terms. Query expansion, on the other hand, is limited to adding new terms to the original query. Although the two terminologies are utilized interchangeably in existing literature, this should not be the case as it refers to specific forms of modification. Thus, in this research effort, the terminology query expansion is used to describe the process of enhancing a query through the addition of related terms.

The query-document vocabulary mismatch problem (Vechtomova, 2009) has been highlighted as the fundamental motivation for query expansion. The main aim of query expansion is to add meaningful terms, phrases or even latent representations to the initial query either manually, automatically or user-assisted. Various query expansion approaches have been proposed over the years. The available methods are grouped as relevance feedback, collection based co-occurrence statistics, thesaurus (ontology) information, according to the source utilized to spawn the additional terms needed for better formulated queries (Chirita et al., 2007). One fairly established method of obtaining additional terms/phrases for query expansion is the relevance feedback method in which users are required to manually suggest/select more vocabulary that may be pertinent to their information need. A variation of this approach would be what is called the pseudo-relevance feedback method, in which the top ranked n documents are assumed to be relevant and from these, terms are selected and used for expanding the query. Relevance feedback methods such as this are not very effective as users are often reluctant to provide explicit feedback, or in general would not wish to prolong the search interaction (Manning et al., 2008, Frakes & Baeza-Yates, 1992). Thus, in order to minimize the search interaction and processing time of a query by avoiding the manual relevance feedback loop, pseudo relevance feedback is adopted where information gathered from statistical processing of term dependencies within a document corpus as well as external knowledge sources e.g. ontology such as WordNet (Available at http://wordnet.princeton.edu/) ) are utilized in the query expansion process. Over the years, a large body of work on pseudo-relevance feedback based query expansion has emerged. Variations of these approaches are mainly in terms of sources used to spawn additional terms i.e. document corpus, external sources (e.g. the WordNet ontology, Wikipedia, search logs) as well as differing retrieval models (e.g. language models, vector space models etc) implemented for the retrieval process. However, an essential factor vaguely addressed in these current works is the linguistic characteristics that depict the intended search goal of user submitted queries.

In this section, background information on the information retrieval process has been presented highlighting the importance of query expansion and the lack of emphasis on linguistic processing in the QE processes. In the remainder, we introduce in Section 2 the main issues impacting the process of query expansion. In Section 3, we provide a module-based decomposition of a generic query expansion framework. For each of the modules, we highlight looming issues and review state-of-the-art solutions in the literature categorized and detailed according to the techniques used. In Section 4, we detail the query pre-processing module that considers the query type according to the user's information need as well as the query structure and length. The linguistic processing module, handling morpho-syntactical and semantic analysis, is presented in Section 5. We then deal in Sections 6 and 7 with the statistical and external knowledge-based processing techniques and in Section 8 with the reconciliation module which consists of term selection and weighting schemes. In Section 9, we develop the novel research directions that can be taken in future efforts and present in particular models integrating statistical and external knowledge-based solutions. Section 10 summarizes and concludes this paper.

## 2   Query Expansion Issues

In word matching based retrieval, the users would have to formulate their queries in the vocabularies of the documents they are trying to obtain. It is a difficult task as large repositories of information would contain different meanings for a particular concept, thus warranting the need for automated QE to minimize query-document vocabulary mismatch (Voorhees, 1994). QE involves deciphering the information need expressed and the pooling of related terms to enhance a user query. In order to understand the intent expressed within a query and to perform beneficial expansion, the identification of key concepts (Bendersky, 2008, Huston, 2010), impact of different structures of a query, its varying length (Avi, 2006; Lau, 2006), and linguistic properties (Moreau, 2007; McCarthy,2003; Morenberg, 2001, Liu, 2005), are important to note.

The study of linguistic characteristics (morphological, syntactical, and semantic) would also aid handling semantic ambiguity (Liu, 2005) that happens in natural language queries, e.g. synonymy and polysemy. In the case of synonymy, a single concept can be represented by multiple words in natural language (as shown in *Scenario 2*). Polysemy is the case wherein single words can have multiple meanings. This ambiguity coupled with the fact that different people express their information need in different ways (as shown in *Scenario 1*) causes poor performance of retrieval.

Query-document vocabulary mismatch can be minimized by recognizing that a query and a document can be either related directly or indirectly through some term relationships. A direct link is when a document and a query contain the same words. An indirect link entails that a document can contain a different word but it may be related to the one in the query (Cao, 2005). Cao and Collins [Cao, 2005; Collins-Thompson, 2005] captured both direct and indirect term relationships through external knowledge sources such as ontologies and statistical processing of the document corpus respectively. Independent usage of the sources showed minimal improvement in retrieval performance. Then, they combined the pooling of both directly and indirectly related terms and showed more improvement in retrieval as compared to individual implementation of the expansion approaches. However, there is only a marginal increase in retrieval performance. The reason behind this minor improvement is due to the fact that several important factors have not been examined and utilized extensively; e.g. the query structure, length, and linguistic characteristics, the term dependency model utilized in statistical processing, and possible terms relationships to extract.

Another important factor that requires significant attention is the reconciliation of the multiple potential expansion terms. Users practice varying QE patterns such as enhancements with synonyms, generalization, specialization, and parallel movement (Rieh, 2006). These practices can be emulated in the reconciliation process to weight the potential expansion terms according to their level of significance within the reformulated query.

We speculate that the reason behind this minor improvement is due to the fact that important factors such as the query structure and length, linguistic characteristics of a query, the term dependency model utilized in statistical processing and possible terms relations to extract have not been examined and utilized extensively.

Another important factor that requires significant attention but not given sufficient emphasis is the fusion of the multiple potential expansion terms. By considering the query expansion patterns that users practice such as enhancement with synonyms, generalization, specialization, and parallel movement (Rieh, 2006), in the fusion process, it would be possible to weigh the potential expansion terms and the category of terms they come from according to their level of significance and importance within the reformulated query. Apart from the factors that have been outlined as being potentially able to enhance retrieval performance, we believe it is worth examining whether the choice of retrieval model plays a role in the effectiveness of the query expansion model as current works show minimal improvement in retrieval performance when embedded within a single retrieval model i.e. language model (Dong et al., 2008; Manning et al., 2008).

## 3   Anatomy of a generic query expansion framework

We propose a generic automated QE framework shown in Figure 5 to duly consider the outlined QE issues via a query pre-processing step and four processing modules:

- The query pre-processing step consists of the identification of the information need encompassed by the query and the determination of query structure and length.

   - The linguistic processing module subjects a query to morpho-syntactical and semantic analyses.
   - The external knowledge base processing module performs sense disambiguation and lexical-semantical related term pooling with external knowledge sources, i.e. ontologies.
   - The statistical processing module performs term dependency modelling for statistically-related term pooling from a document corpus
   - The reconciliation module deals with the integration of potential expansion terms derived from linguistic, knowledge-based and statistical sources through term mapping and weighting strategies.

At the core of this framework are the significant constituents that characterize the query intent. Their identification within a query is crucial to determine only terms which directly express the information need of the user and are not merely structural components of a query. They are isolated through two processes within the Linguistic Processing module (detailed in Section 5), i.e. the non-compositional phrase isolation and syntactical analysis. Their processing is done at two levels, local and global. Local analysis requires that they are subjected to the external knowledge-based processing module (Section 6) and the statistical processing module (Section 7) to obtain potential expansion terms and be assigned suitable weightage on their own accord. On the global level, relationships between them that are inferred through linguistic analysis and statistical processing would be given due consideration in the reconciliation and expansion module (Section 8) and appropriate weights assigned to the constituents characterizing the intent within the context of the query.

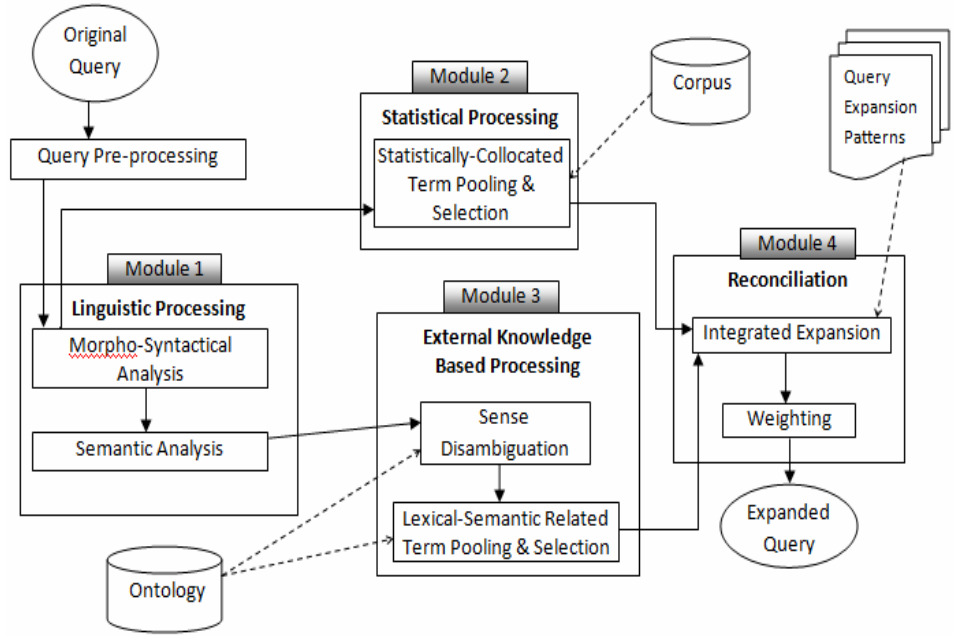

**Figure 5 Overall Organization of a Generic Query Expansion Framework**

## 4. Query Pre-processing

### 4.1. Information Need & Query Type

An essential aspect for consideration when dealing with information retrieval is the information need expressed by users. Search goals can be broadly categorized as exploratory, fact-finding, and comprehensive (Aula, 2003). In exploratory search, the aim is to get a general idea of a topic where high recall or precision is not required but rather having a few documents as a reference is sufficient. In fact-finding search, the precision of the results returned is very important as compared to comprehensive search in which high recall is of greater importance. Broder (2002), Rose & Levinson (2004) and Jansen et al. (2008) concur with several previous studies, that queries of type informational, navigational or transactional are employed to express exploratory, fact-finding and comprehensive search goals. Informational queries are formulated with the intent of locating content (e.g. data, texts, documents or multimedia) about a particular topic. (Example: "help quitting smoking"). Navigational queries are expressed with the intent of locating a particular website of a person or organization (Example: "St. Joseph's hospital"). Transactional queries represent a goal of obtaining a product through the act of purchasing, downloading or executing (Example: "Josh Groban's song"). Among the three query types, informational queries are most widely issued by users and account for approximately 60% of submitted queries, transactional queries at 27%, navigational 12% and the remaining 1% of queries could not be classified. An information need, whether exploratory, comprehensive or fact-finding, is expressed in the form of informational, navigational and transactional queries. However, given that these are natural language queries, they may vary in terms of their structure and length. Users may formulate natural language queries which might be grammatically correct or incorrect and may vary in terms of its length.

### 4.2. Query Structure & Length

Informational, navigational or transactional queries submitted by users are structured in a variety of ways. We note from a preliminary observation of query lists obtained from "website optimizing sites" such as Google Analytics (Available at http://www.google.com/analytics/), Stat Counter (Available at http://www.statcounter.com/)and also the TREC 2007 Million Query Track (Available at http://trec.nist.gov/data/million.query07.html) that a user provided query may take the form of bag-of-words or well-formed sentences. Well-formed sentences are typically grammatically correct but may also be partially complete sentence fragments. On the other hand, bag-of-words queries are simply a list of concepts which would typically relate to one particular topic.

Apart from the structure of queries, several studies have been conducted to examine the length of queries submitted by users. Lau et al (Lau & Goh, 2006) observed that approximately 75% of all submitted queries constituted not more than 3 terms and only 25% were queries with more than 3 terms. Several other studies (Arampatzis & Kamps, 2008; Wolfram, 2008; Phan et al., 2007; White & Morris, 2007) reported similar findings wherein average length of queries vary from 2 to 4 terms with the longest of queries being between 18 to 49 terms. Phan et al. (2007) attempted to find the correlation between length of queries and the degree specificity of user information need. The average cross-over point of specificity between broad and narrow queries is observed at 3 terms, from the range of 1-10 terms per query. It is concluded that users tend to use short queries in general and as query length increases, the corresponding information need is more likely to be perceived to be narrow. Voorhees (1994) highlights that long queries, if not well detailed, would also benefit from query expansion as much as short queries. This is similar to Lau & Goh (2006) who theorize that as query length increases, there is a higher probability that users would encounter unsuccessful searches. However, long queries may not always have done well with direct query expansion as such queries may consist of multiple important concepts. A verbose query may not always contain explicit information in the description itself to indicate which of these concepts is more important. Secondly, such queries may contain two or more equally essential key concepts. We believe it is thus vital to identify the key concepts expressed in a query to ensure the main search intent is emphasized.

### 4.3. Analysis & Open Issues

Two important properties of a query that have been highlighted are the structure and length of a query. There is a need to investigate whether the different structures of a query, bag-of-words or sentences, and its varying length would have an impact on the query expansion strategies. For both, bag-of-words and sentence type queries, adjacent terms within a query may provide context information that may enable the sense of a term to be accurately disambiguated. Further, syntactical relations between pairs-of-words in a sentence-type query might be scrutinized to capture the intended information need. Multiple concepts that occur in especially long queries should also be identified and possibly ranked in order of importance to ensure the main goal of the query is captured. For this purpose, a query should be analyzed based on its linguistic characteristics.

## 5. Module 1: Linguistic Processing of Queries

In order to decipher the exact intent of a query, queries may be scrutinized based on their linguistic characteristics. However, a full study of natural language query characteristics in terms of morphological, syntactical and semantic properties can be a challenging task as there is much depth involved. Thus, through a survey of computational linguistics studies, three major linguistic properties of queries (i.e. morphological, syntactical and semantics) that can be extracted for the purpose of query expansion are identified.

### 5.1 Morphological processing

Morphology is the identification, analysis and discription of the structure of words. Several morphological properties of natural language are identified through Goldsmith's work (2001) in which the process of unsupervised learning of the morphological segmentation of words into the components that form the word by the operation of concatenation is discussed. The author notes that inflections and derivation of words in natural language occur mainly from prefixation and suffixation. Inflection simply means that there are variant forms of the "same" word (e.g. dog, dogs) whilst derivation is a form of word-formation where the addition of an affix derives a new word (e.g. dishwasher). Compounding is another form of word-formation where it involves combining complete word-forms into a single compound form (new form) e.g. dog-catcher. Moreau et al. (2007) highlights that morphological variants have an impact on retrieval performance and propose an unsupervised machine learning technique for morphological analysis. The fundamental idea is to detect terms for query expansion, within a collection of texts, which have morphological relations with the terms of the query. These relations are identified through an analogy learning scheme that banks on a similarity notion that is used to determine whether two pairs of words are analogous (i.e. when two pairs of words have a high degree of relational similarity). The similarity measure attempts to capture inflections and derivation within a language, which mainly is obtained through prefixation and suffixation. The results of their experiments show that statistically significant improvement can be achieved when an average of three morphological variants are added to each query term, regardless of query length and morphological process.

Two common approaches of morphological analysis of words are stemming and lemmatization. The two differ in that stemming most commonly collapses derivationally related words, whereas lemmatization only collapses the different inflectional forms of a word. Both approaches (stemming and lemmatization) can be utilized especially since the idea of query expansion is to include terms that may be related or useful for accurate retrieval. However, stemming is widely utilized in information retrieval techniques as compared to lemmatization. This is because lemmatization (despite being more accurate than stemming) is a computationally intensive process compared to stemming. Peng et al. (2007) claim that traditional stemming which entails the transformation of words into their root form before indexing and query expansion, increases recall but lowers precision for Web Search, apart from incurring additional computation cost. Thus, the authors propose context-sensitive analysis of the query to avoid blind transformation of all query terms without considering the context of the word. The authors limit stemming to handling only pluralisation. Candidate terms are generated offline using the Porter Stemmer method that converts a word to its base form and also by conducting a corpus analysis based on the word distributional similarity in which the authors assume that true variants tend to be used in similar contexts. Among their experimental results, one observation is most significant in that blindly pluralizing short queries is relatively safe. However for long queries, most queries can have multiple words that can be pluralized. Expanding all of them without selection will significantly hurt precision. The authors highlight the possible errors that could occur through pluralisation and de-pluralisation due to the fact that they are not symmetric. From the linguistics point of view, a plural word used in a query indicates a special intent and thus de-pluralisation may bring specificity to the original information need (e.g. London hotels -> London hotel).

## 5.2. Syntactical processing

Syntactical analysis is the study of the principles and rules applied in sentence construction of natural languages. A query could consist of compositional and non-compositional phrases. A compositional phrase in a query is made up of its individual components or terms whereas in the case of non-compositional phrases (e.g. idioms, proper names, collocations, acronyms, phrasal verbs, modals, fixed phrases discourse markers) a combination of all its component terms represent a larger meaning e.g. red tape (Lin, 1999).

Syntactic parsing of natural language is crucial for the discovery of constituent representations and grammatical relations of sentence (McCarthy & Carroll, 2003; Morenberg, 2001; Punyakanok et al., 2008). Syntactical analysis is particularly necessary when dealing with compositional phrases and this encompasses the rules that govern the combination of constituent representations (e.g. POS e.g. noun, verb etc; phrases e.g. noun phrase, verb phrase etc) and grammatical relations (typed dependency relations between pairs of words e.g. subject-object, verb-object) in forming a sentence to express a particular intent. Constituent representation of a query is the distinction of its portions as noun (NP), verb (VP), preposition (PP), adjective (ADJP) or adverbial phrase (ADVP) whilst POS is the labelling of query terms as one of the eight POS within the English Language, i.e. noun (N), pronoun (PN), verb (V), adverb (ADV), adjective (ADJ), preposition (P), conjunction (C) and interjection (I).

Gao et al. (Gao et al., 2004) propose a query expansion framework based on linkage structures that take the form of acyclic, planar, undirected graphs for capturing syntactical dependencies in a sentence. After a first step of linguistic parsing, the computation of these structures is achieved through unsupervised learning on example sentences. Expansion terms are then generated based on the most important syntactical dependencies. Lioma & Ounis (2008) present a query expansion technique that is based on syntactical analysis. The authors attempt two approaches: firstly, a purely shallow syntactic-based query expansion (SSQE) technique and second, a combination of the SSQE method and the probabilistic pseudo relevance feedback approach. The fundamental idea of SSQE is to reduce the query size by retaining only sentential fragments that correspond to the POS blocks assumed to be content-rich, on the basis of their high frequency. In their approach, further noise in a query is eliminated through the removal of periphrastic structures i.e. phrases introducing a point, rather than the factual point itself. The core idea of this approach assumes that the relation between the frequency of occurrence of POS blocks in a representative language sample and the amount of content they bear is approximately directly proportional. This is not an accurate assumption as frequently occurring POS blocks are merely a result of sentence construction in natural language documents. Apart from that, this approach is highly computationally intensive as it requires that every document in the collection is parsed to identify its POS blocks. A probabilistic retrieval model is utilized in this framework.

### 5.3 Semantic ambiguity

Semantic ambiguity is a long recognized problem in natural language processing [Efthimiadis, 1996). Multiple words may share the same meaning (synonymy). For example, the term automobile and vehicle both refer to a mode of transport. Another form of semantic ambiguity is the case where a single word may have multiple meanings (polysemy). For example, the term "book" may refer to a source of knowledge or the act of reserving something e.g. a flight ticket. A first level of disambiguation may be achieved through syntactic parsing where the partofspeech (POS) of a term is determined based on the context of its usage within a sentence (Stevenson & Wilks, 2001). As such, recognition of POS and syntactical dependencies, such as proposed in the query expansion frameworks of Gao et al. (2004) and Lioma & Ounis (2008), not only assists in deciphering intent within the context of a query but also contributes in sense disambiguation of ambiguous terms. Once a POS is assigned, a single word sense could be selected through a process of inference which depends on measures of semantic relatedness or semantic distance among query terms (Budanitsky & Hirst, 2006; Turney, 2006). Semantic measures are obtained through an external knowledge base, e.g. the WordNet ontology , and sense disambiguation is therefore performed in the second module handling external knowledge-base processing.

### 5.4 Analysis & Open Issues

We note that morphological variants, when carefully selected with regards to their context of usage, enhance retrieval performance. Even so, the level of importance given to morphological variants is probably of less significance as compared to synonymous terms and semantically related terms in addressing the issue of synonymy and polysemy. Each of the eight POS (i.e. verbs, nouns, pronouns, adverbs, adjectives, prepositions, conjunctions, and interjections) play its own role within a query whether it be representing a name, an action or merely joining two words. Scrutiny of a query from the linguistic perspective wherein syntactical dependencies are captured that eventually leads to the identification of a head noun. A "head noun" is a concept upon which everything in a query is centered. This could be of help in deciphering the most important concepts of a query, especially in the case of long queries. Identification of POS of a query provides another mode of sense disambiguation and iterates the role of a term within a query. A query is made up of a series of terms that are grouped together in a sentence or sentence fragment to express a particular intent. However, queries are not always formulated as complete sentence or even a sentence fragment. A sentence fragment can be crudely differentiated from a bag-of-words as being a partially complete sentence which consists of one type of phrase (e.g. noun phrase, verb phrase, adjectival phrase etc). A bag-of-words, on the other hand, if analyzed from a strict linguistic point of view, would be defined as a group of words with absolutely no syntactical relations between the terms. Nonetheless, semantically, adjacent terms in a query could provide context to neighbouring ambiguous words and thus, shed some light about the actual intent of a query. With this in mind, the definition of a bag-of-words type of query needs to be redefined to ensure that every bit of useful information is extracted in understanding a query.

A study of the linguistic characteristics of a query allows for a more precise and structured approach in deciphering the intent of a query and its most significant concepts that form the subject of the information need. Recognizing that linguistic processing may be error-prone due semantic ambiguities that occur in natural language queries, current works (e.g Bendersky et.al,2009, Huston & Croft, 2010) incorporate a combination of syntactical analysis and statistical mining in key concept identification tasks. Also, some query expansion techniques (e.g. Liu et al, 2005) incorporate incorporate sense disambiguation techniques (e.g. Pedersen et al, 2004) to accurately decipher the intended search goal of queries.

### 6 Module 2:   External Knowledge-Base Query Processing

External knowledge-based query processing relies on external sources (such as ontologies, search logs, web corpus (e.g. Wikipedia) etc ) for obtaining expansion terms. We discuss in Section 6.1 the use of knowledge bases in handling the semantic ambiguity issue introduced in Section 5.3.

Another issue, dealt with in Section 6.2, is linked to the highlighting of term relationships captured by the external knowledge bases. We then detail techniques for knowledge-based processing query expansion models in Section 6.3.

## 6.1 Sense Disambiguation

Among the approaches conducting word sense disambiguation, Kim et al. (2004) aim to improve the performance of large-scale text retrieval. A root sentence tagger classifies each noun in the documents and queries into one of the 25 root senses found in WordNet 2.0, which is called coarse-grained disambiguation. When classifying a given ambiguous word into one of the root senses, the most informative neighbouring clue word having the highest mutual information with the given word is first selected. Then, the single most probable sense among the candidate root senses for the given word is chosen according to the mutual information between the selected neighbouring clue word and each candidate root sense for the given word. Considering only the single most informative neighbouring word in a corpus as evidence of determining the sense of a target word is a good approach to narrow down multiple variations in senses. However, the authors fail to consider verb senses in their approach. This may not be wise as verb senses could bring further improvement in understanding the intended meaning of the query.

Liu et al. (2005) perform word sense disambiguation in short queries, choosing the right sense for a word in its occurring context. Noun phrases of the query are determined and compared against synonyms and hyponyms from WordNet to identify word senses. If it is not possible to determine the word sense in this manner, then a guess is made based on the associated frequency of use available in WordNet. Frequency of use is determined by the number of times a sense is tagged in various semantic concordances. The approach of utilizing the "first sense" heuristic (choosing the first, or predominant sense of a word) is reiterated by McCarthy et al. (2007) as a method that is used by most state-of-the-art systems as a back-off method when text-based disambiguation is unsuccessful.

## 6.2 Eliciting Term Relationships for Related Term Pooling

Recognition of term relationships would enable a more systematic approach to term selection for query expansion. Term relationships can be categorized as lexical and semantic relations. Lexical relations are defined as relations between basic word forms that share the same meaning such as synonyms and morphological variants. Semantic relations, on the other hand, represent relations between meaning of words such as synonymy, hyponymy/hypernymy (IS-A) and meronymy/holonymy (HAS-A) (Miller et al., 1990). Apart from these formally defined relations, pairs of words may also be related based on their positional dependency within a given proximity in a document. Such word pairs are said to be related to the same topic if they frequently co-occur within a document corpus. Extraction of both formal and informal term relationships would ensure that all relevant terms to a query topic are considered during expansion.

A query and a document can be either related directly or indirectly through some term relationships. A direct link is when a document and query contain the same words whilst an indirect link entails that a document can contain a different word but it may be related to the one in the query, either through synonymous words or other relations (Cao et al., 2005). Query expansion patterns identified by Rieh & Xie (2006) require that specific type of terms be utilized to achieve the four types of content modification i.e. replacement with synonym (synonyms), generalization (hyponyms/hypernyms), specialization (meronymy/holonymy) and parallel movement (coordinate terms). These lexically and semantically related terms can be derived from the WordNet ontology ("WordNet", 2006). Statistical processing of the document corpus would be required to extract relevant frequently co-occurring terms that represent proximity-based relations. The utilization of both the formally and informally related terms in query expansion, coupled with other factors discussed above, is anticipated to minimize the query-document vocabulary mismatch problem.

### 6.3. External Knowledge-Based Processing Query Expansion Models

State-of-the-art frameworks rely on ontologies as external knowledge-based sources to obtain term suggestions. Ontologies may be domain specific or general (Mena et al., 2000). Query expansion using domain-specific ontologies are more suitable for static document collections. For web collections, the ontologies would have to be frequently updated because the collections on the web are more dynamic in nature.

In Figure 6, We illustrate the general process flow of current ontology-based query expansion frameworks. A query is subjected to three major processes i.e. linguistic processing, sense disambiguation and lexical-semantic term pooling. An original query ("how to repair a car") is subjected to stemming and stop word removal. The remaining concepts in the query are then disambiguated against an ontology using a semantic measure of relatedness (e.g. Pedersen et al, 2004). Based on the derived sense of the query concepts, lexical and semantically related expansion concepts are extracted from an ontology. Terminologies in domain-specific ontologies are less ambiguous therefore short queries can be expanded with a higher chance of accuracy. General ontologies would be suitable for broad queries (Bhogal et al., 2007). In previous research efforts involving external knowledge sources, Voorhees (1994) and Liu et al. (2004) utilize general ontologies while Radhouani et al. (2006) use domain specific ontologies.

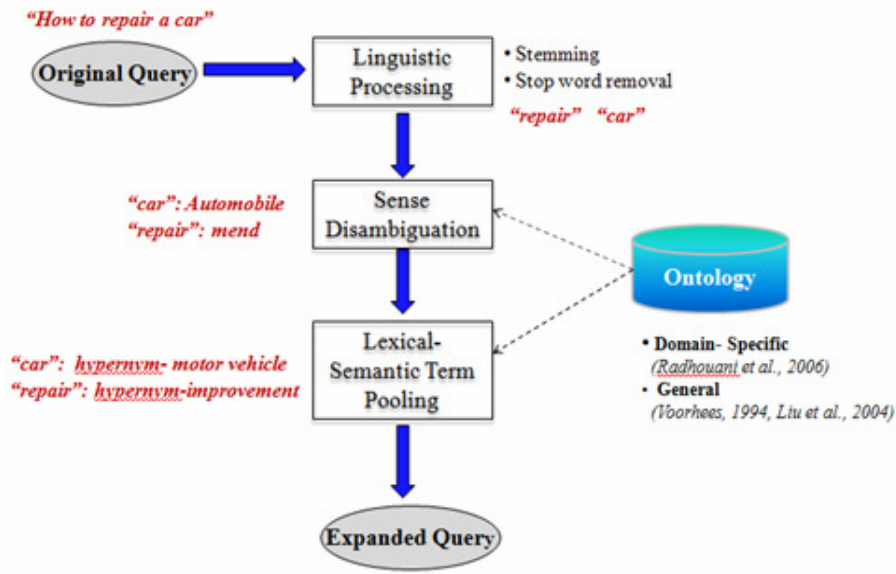

**Figure 6: Overview of External Knowledge-Based Query Expansion Models**

### 6.4 Analysis & Open Issue

Domain specific and general ontologies have been utilized for the purpose of extracting semantically related terms and sense disambiguation (Liu et al., 2004; Voorhees, 1994), phrase extraction (Liu et al., 2004) and query dimension identification (Radhouani et al., 2006) in the external knowledge-based query expansion efforts discussed above. Query structures and varying query lengths were not differentiated during the process of expansion in these efforts. However, retrieval performance in long queries may be improved by modelling the work done by Rhadouani et al. (2006). In their work, the semantic dimensions of medical queries were identified and a document is considered relevant if it contains one or more dimensions of the query. Unfortunately, identification of dimensions of a query is not straight-forward in general queries as the terms utilized in such queries and those found in general ontologies are highly ambiguous. For non-domain specific queries, this approach of identifying dimensions could be likened to the identification of main query concepts. Main query concepts can be identified in long queries, especially, allowing documents which contain one or more of the main concepts to be deemed as relevant. Queries are subjected to minimal linguistic analysis where they are stemmed and stop words removed prior to the process of expansion. From the syntactical analysis point of view, Liu et al. (2004) recognized the importance of identifying phrases that occur in a query instead of treating all terms in a query independently. Liu et al. rely on an ontology to identify proper names and dictionary phrases but it is not clear whether they cover all crucial non-compositional phrases such as phrasal verbs, idioms, collocations, acronyms etc. Also, they do not utilize syntactical dependencies to identify simple and complex phrases that are not found in the dictionary but instead devise a simple grammar that relies on the existence of nouns phrases and content words (non-stop words) within the query to recognize phrases. Identification of non–dictionary phrases could very well be improved by instead relying on Google N-gram ("Google N-gram", 2006) frequency counts that lists commonly found phrases in web pages. Both Voorhees (1994) and Liu et al. (2004) utilize the ontology to perform semantic analysis e.g. word sense disambiguation to address the polysemy issue. Liu et al's approach was more extensive than Voorhees and worthy of further exploration as it considers adjacent query terms to provide contextual information for sense disambiguation. Query expansion patterns were not distinguished in any of these research efforts. Voorhees (1994) and Liu et al. (2004) utilized only one source of potential expansion terms that is the external knowledge base. Both works include lexically related terms, synonyms and hyponyms. Voorhees also utilized hypernyms and meronyms whilst Liu et al. utilized compound terms as well. This mix of concepts may not work well when terms which are of contrasting relations to a query topic (e.g. generalization and specialization) are utilized together causing the range of related documents to be less precise. We believe that the results obtained from these attempts can be improved by considering the expansion patterns. Apart from that, proximity based related terms were not considered in the expansion process. Radhouani et al. (2006) utilize the ontology to identify the dimensions of an original query but not for extraction of related terms. Voorhees (1994) and Radhouani et al. (2006) evaluated the expansion model within the vector space model whilst Liu et al. (2004) utilized a probabilistic model. All the research efforts discussed do not examine the effectiveness of their approach on multiple retrieval models.

The effectiveness of query expansion via external sources is very much dependent on the accurate disambiguation of query concepts. However, sense disambiguation methods are far from perfection resulting in multiple research works that attempt to improve the accuracy of disambiguation through improved measures of relatedness (e.g. Patwardhan,2007) and via information extracted from external sources such as Wikipedia (e.g. Mihalcea, 2007).

A noticeable limitation of using the Wordnet Ontology for query expansion is the limited coverage of concepts and phrases within the ontology. Query expansion via other external sources such as Wikipedia (e.g. Yinghao, L., et al., 2007) and search logs (e.g. Chu,H. et al, 2003) are of particular interest to fill the gaps of the Wordnet Ontology, if not replace it. It is our postulation that external knowledge-based query expansion would benefit most in the case where multiple external sources are utilized in the expansion concept extraction process. This is because ontologies such as WordNet provide semantically-related expansion concepts whilst other corpus-based external sources rely on concept proximity to determine relatedness of concepts. This restricts our ability to ensure that both direct and indirect links to a document are captured in the process of query expansion (described in Section 6.2)

## 7   Module 3: Statistical-Based Query Processing

Statistical processing relies on a document corpus for obtaining expansion terms. In Section 7.1, we discuss term dependency modelling for pooling statistically collocated expansion terms. We then discuss current statistical processing based query expansion models in Section 7.2.

## 7.1 Term Dependency Modeling

Several statistical processing approaches exist that model term dependency for the extraction of proximity based terms (Bai et al., 2005; Gao et al., 2004; Lioma & Ounis, 2008; Metzler & Croft, 2005, 2007; Song et al., 2008). Classical information retrieval models usually assume independence between terms which entails queries and documents to be treated as bags-of-words. This assumption is disagreed upon in recent times by researchers who acknowledge that there exist dependencies between terms and if considered appropriately would improve retrieval performance (Song et al., 2008). Furthermore, experimental results in (Metzler & Croft 2005) show that modelling dependencies can significantly improve retrieval effectiveness across a range of collections. Term dependencies are present between terms within a query or between terms within a document as well as between query and document terms. Typically, one observes the frequency of term co-occurrences within a certain context, which can be the whole document, passage, or a window of fixed length. Then the strength (or probability) of term relationship is computed. Initial attempts at modelling dependencies considered individual terms which were assumed to be independent of each other. Bigram and bi-term models were then introduced to include adjacent dependencies that occur between pairs of words that occur together. However, as pointed out by Gao et al. (2004), adjacent terms which are not truly connected ("noise") would also be considered in the bigram model, which in turn allows for only a marginal increase over the performance of the unigram model. Term dependencies do not only occur between adjacent terms but also between terms which are positioned within a larger distance of each other. Such distant dependencies cannot be correctly covered by a bigram or bi-term model, therefore further analysis has been carried out in recent research efforts and its impact on query expansion is discussed in the following section.

## 7.2 Statistical Processing-Based Query Expansion Models

Two main statistical term dependency models instantiated in query expansion frameworks can be highlighted in the literature: the first captures dependent terms through the computation of their frequency of co-occurrence in a document corpus (Metzler & Croft 2005, 2007) while the second considers the distance between these terms (Gao et al. 2004; Song et al. 2008). In Figure 7, we illustrate the process flow of recent works that model term dependency in statistical processing-based query expansion. An original query is typically subjected to fundamental linguistic processing of stemming and stop-word removal prior to processing term dependencies. An expanded query is formed depending on the chosen dependence model e.g. when full independence among query terms is assumed, expansion concepts extracted are e.g troubleshooting, sales and review; whilst with sequential dependence applied, expansion concepts extracted are e.g. workshop and breakdown.

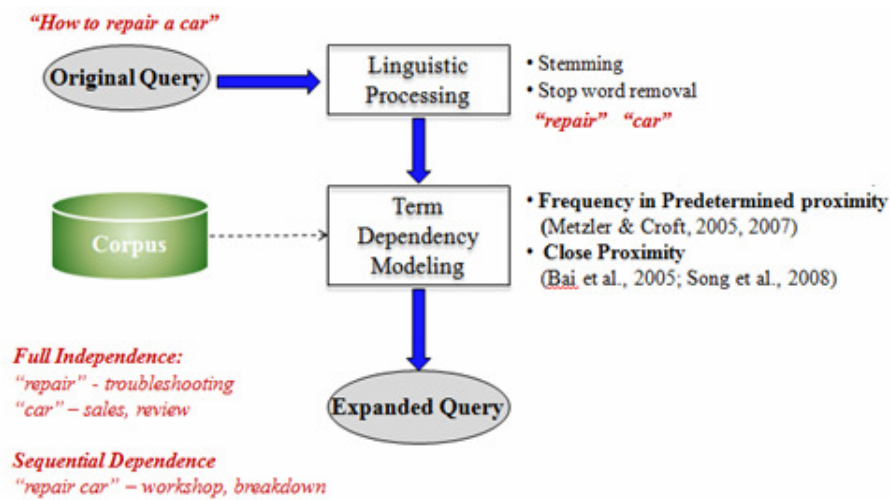

**Figure 7: Overview of Statistical Processing-Based Query Expansion Model**

In the first approach, three variants of term dependency modelling can be considered, i.e. full independence, sequential and full dependence. Full independence assumes that all query terms are independent, whilst the full dependence variant captures dependencies that exist between every subset of query terms. Sequential dependence assumes that certain dependencies exist between adjacent query terms. Experimental results in (Metzler & Croft 2005) show that the sequential dependence variant is more effective on smaller, homogeneous collections with longer queries, whereas the full dependence variant is best for larger, less homogeneous collections with shorter queries. This dependence model is instantiated within a query expansion framework in (Metzler & Croft 2007). Through a technique coined Latent Concept Expansion (LCE), the authors propose single or multi-term expansion by modelling term dependencies and the use of arbitrary features such as named entities, term proximity, text style, document length, PageRank, readability etc. No explicit model of syntax or semantics is used; rather, the authors use pure term co-occurrence in the query and feedback documents. From the experimental results, it is evident that the LCE technique improves effectiveness for 65%-80% of queries, depending on the data set. The retrieval model utilized in this model is the language model. It is useful to note that while LCE can perform multi-term expansion, the authors report that improvements are for single-term expansion only.

In the second approach, Song et al. (2008) model dependency wherein query terms are grouped into non-overlapping phrases according to the distance between them within a corpus. The authors try to solve two problems that occur in previous dependency studies: the difficulty to estimate the importance of phrases because they are different in nature from words; and as phrases are not independent from the component words, a linear combination of scores for words and phrases is inappropriate. The relevance contribution of a component query term occurrence is measured by how many query terms occur in the context phrase and how far apart they are within the phrases. Finally, the relevance contribution from each query term is accumulated. By replacing term frequency with this accumulated relevance contribution, term proximity is incorporated into a probabilistic retrieval model.

A combined approach is proposed in (Bai et al., 2005) where the authors perform query expansion based on the Information Flow method which extract inferential term relationships. The proposed technique is similar to Metzler & Croft's (2007) in terms of the utilization of co-occurrence statistics within a scope of proximity but the measure of association between two terms is similar to Song et al's approach where it is based on the distance between two terms within a predetermined window in a document. Unlike the traditionally used pair-wise term relationships, these relationships are context-dependent, in the sense that they are between a set of terms and another term. The frequency of term co-occurrences is observed within a certain context, which can be the whole document, passage, or a window of fixed length. All the words occurring within the contextual information are considered as co-occurring with each other. The strength of association between two words is inversely proportional to their distance. The retrieval model utilized in this framework is the language model.

### 7.3 Analysis & Open Issues

In the statistical processing approaches that allow for the extraction of proximity based related terms, query structures and varying query lengths are not differentiated during the process of expansion. Queries are subjected to minimal linguistic analysis where they are stemmed and stop words removed prior to expansion. We suggest that subjecting only the most meaningful concepts in a query (derived from linguistic analysis) that have been disambiguated, to statistical processing would ensure that only query concepts that truly represent the query intent are expanded. Query expansion patterns were not distinguished in any of these research efforts. Informally defined term relationships which are the proximity based related terms were extracted whilst formally defined relations from external knowledge sources were not considered in the expansion process. Bai et al. (2005), Gao et al. (2004), Lioma & Ounis (2008), Metzler & Croft (2007) and Song et al. (2008) utilize the probabilistic retrieval model and do not examine the effectiveness of their approach on multiple retrieval models.

A question arises on how useful it is to analyze term dependencies in queries for the purpose of information retrieval. Lang et al (Lang et al, 2008) conducted a study on the significance of incorporating term-dependencies in information retrieval tasks. The authors hypothesize that not all queries may benefit from the integration of dependency estimation. A model was constructed to calculate term statistics and term dependency of a query. Term statistics is computed by calculating the average term frequency of query terms for each document in the relevant and irrelevant documents. On the other hand, term dependency statistics is computed by calculating the average occurrences of ordered phrases in a document instead. From this study, the authors found that incorporating term dependency can improve the average retrieval performance, in which 59% of the queries benefit from the incorporation of term- dependency, whilst the remaining 41% does not. The authors conclude since not all queries would benefit from considering term-dependencies, it would be good to identify queries for which the dependency model would work. However, the difference in mean average precision with and without identifying queries that would benefit from term dependency modelling is minimal. Thus, there may not be a necessity to predict whether a query would benefit from the dependency model, especially since the proposed prediction model is computationally intensive. Also, the reason that some queries do not perform well could be due to other reasons such as the linguistic properties of the query and not directly due to the dependence model. It is also acknowledged that unsupervised statistical methods may have restricted ability to capture due to its consideration of term proximity and frequency of co-occurrence rather than semantic associations/relatedness among query concepts in the expansion term extraction process.

## 8. Module 4: Reconciliation & Query Expansion

Direct and indirect relationships between queries and documents require that formally and informally defined term relationships be extracted and incorporated. These potential expansion terms would be extracted from multiple sources such as a linguistic parser, the document corpus and ontology. To emulate user tendencies when reformulating their queries to obtain desired results, we think it is necessary to distinguish the query expansion patterns practiced by users. Having multiple terms added into the original query, it is then a challenge to devise a term weighting scheme that would adequately signify the role of the query terms in representing the query intent.

### 8.1 Term selection and Query Expansion Patterns

The large percentage of informational queries issued representing fact-finding and comprehensive search goals warrants the need to aim for striking a balance between high levels of precision and recall. Maximizing precision requires that queries contain precise terms whilst maximizing recall requires the use of broader and more general terms. Novice users of search systems tend to formulate broad queries whilst a minority of advanced users tend to be more precise in their search (Aula, 2003). A high proportion of users do not utilize advanced search features and those who do frequently misunderstand them (Lau & Goh, 2006). The authors observe that the accuracy of query formulation depends on a user's familiarity with search tasks and of the subject matter. Users are plagued with the inability to accurately represent their information needs due to a number of factors which are, among other things, lack of familiarity with domain specific vocabulary, lack of clarity in expressing their intent, lack of familiarity with the search environment and minimal awareness of the logic applied on to their query when a search request is made (Salton & Buckley, 1990; Spink et al., 1998; White & Morris, 2007). Consequently, the returned search results are very often not to the users' satisfaction which then leads to the expansion of their initial query.

Rieh & Xie (2006) examined query logs and observed that typically query expansion attempts are modifications of content, format and resource facets.
Content modification refers to changes made to the meaning of a query inclusive of:

- Generalization where users generalize the meaning of the query by deleting terms or replacing terms with those that have more general meaning (e.g. "best buy San Antonio" being modified to "best buy store") and
- Specialization in which users specify the meaning of the query by adding more terms or replacing terms with those that have more specific meaning (e.g. "lose job" being modified to "lose job effects") of a query
- Replacement with synonyms where Users replace current terms with words that share similar meaning (e.g. "Dayton equipment" being modified to "Dayton tools")
- Parallel movement where users modify queries to have partial overlap in meaning (e.g. "America West airlines" being modified to "Delta airlines")

Format modifications include the cases in which users made changes without altering the meaning of the query by means of term variations (e.g. "FIT" being modified to "Fashion Institute of Technology"), operator usage (e.g. "trajectory methods" being modified to "trajectory and methods") and error correction (e.g. "dayton equipmen" being modified to "dayton equipment"). Resource modifications are those instances in which users intended to make changes in types of information resources (e.g. "news on Iraq war" being modified to "images of Iraq war") and domain suffixes (e.g. "camelot" being modified to "camelot.com").

Among these categories, content modifications account for 80.3% of query expansions, 14.4% of the query expansions are related to format alone and only 2.8% of the modifications are associated with resource expansion whilst the remaining 2.5% of the expansions cannot be defined. Given its small percentage of occurrence and there being no basis for deciding to change the type of resource requested unless prompted by the user, resource modifications would not be handled in all automated query expansion approaches. On the other hand, a conservative amount of format modification would take place during the expansion process whilst content modifications were the most inherent form of query expansion. The results of Rieh & Xie's study (2006) reveal that users do not always start with a general query and attempt to specify it. In fact, parallel movement is the most popular means of content modification (51.4%) while specification accounts for 29.1% and generalization 15.8%. Users do not often change a query by simply replacing it with a synonym: only 4.9% of all expansions are related to patterns of replacement with a synonym. Among the four forms of content modification, the most directly applicable expansion is the replacement with a synonym modification.

Another approach of query expansion is proposed by Bozzon et al. (2007) in which POS pattern transition in query expansions are analyzed. The authors note that most queries are composed of only nouns (33%). The authors note that a large portion of these terms are ambiguous and find that such queries are reformulated to contain noun-verb (multi-sense nouns) combinations. Very rarely there would be queries composed of adjectives and/or adverbs and these queries are typically not reformulated. There is a significant amount of queries that contain four out of the eight available POS: (noun (N), verb (V), adjective (Adj), adverb (Adv)). The authors highlight that the major transition patterns from these queries involve either adding even more words, or removing the adjective(s). Also, the authors claim that users generally further extend queries of the form N-V-Adv in an attempt to specialize the query whilst users perceive the N-V-Adj as being too specific and as a result remove terms from it. We believe that applications such as automatic query expansion could be improved by knowing which POS are more likely to be added or removed by each user. On the contrary, these findings only reiterate the findings of other user studies whereby users generally add more words to their queries when they reformulate queries. However, information about the POS of queries that users add or remove to narrow or broaden queries would not be as useful in selecting terms for expansion as compared to the type of terms that they would typically add (e.g. terms that specialize or generalize a query) when they reformulate a query.

An interesting observation by Bozzon et al. (2007) is that a large portion of queries are composed of only nouns but are generally reformulated to contain noun-verb combinations. An interpretation of this phenomenon is that nouns and verbs are a central part of a query and should not be dropped in the automated query expansion process. Unlike nouns and verbs, adjectives and adverbs are dropped when users resubmit their queries. This is possibly due to the role of adjectives and adverbs that serve to describe nouns and verbs further but do not play a significant role in a query individually. However, this may not necessarily be the case especially when dealing with multimedia retrieval queries in which adjectives play an important role to accurately describe an image or video. As an example, in describing an image that contains a blue car parked in front of a shopping mall, the adjective blue is essential in ensuring that images that contain only the representation of a blue car are retrieved. More specifically, low-level features such as color, texture and shape of multimedia objects are expressed through the use of adjectives (Hollink et al., 2004). The authors note that users tend to add more words to their initial query when reformulating it. The type of words added to a query differs based on the form of query expansion practiced by the user. Rieh & Xie (2006) have highlighted content modification comprising of generalization, specialization, synonym replacement and parallel movement as the most popular form of query expansion. Among the four, users showed the most preference towards parallel movement and the least preference towards replacement with synonyms (Mastora et al., 2008; Rieh & Xie, 2006). Instead of replacing synonyms, we suggest that enhancement of queries with synonymous terms would ensure that most directly relevant documents are retrieved.

It is, however, not as clear-cut to incorporate parallel movement, generalization and specialization in automated expansion systems as some amount of user feedback would be required to avoid over-generalizing or over-specializing the original query. A possibility that may be explored is to provide the users with results of queries that are reformulated through parallel movement, generalization and specialization respectively. This would provide users with all possible documents related to their initial query, thus eliminating the need to reformulate queries repeatedly to view various contents. We propose further investigation in the form of user surveys to derive rules that may be utilized to predict the desired query expansion pattern depending on the original categorization of a query type (described in Section 4).

Mastora et al. (2008) also note from their study that 74.5% of users used a term from the initial retrieved results when they reformulate their query. The authors do not specify where these terms are taken from, i.e. the document title, body or snippet. This shows that documents from the corpus containing relevant original query terms is a good source of more relevant terms that may be utilized for query expansion. This further motivates the implementation of integrated query expansion frameworks to adequately capture both direct and indirect query document links.

**8.2 Information Retrieval Models and Term Weighting**

Information retrieval models are utilized to support the processes of indexing and retrieval. Several different models of information retrieval have been introduced by the information retrieval research community. There are three distinct types of models namely the set-theoretic, algebraic and probabilistic model, which differ in the way the document and query terms are organized and estimated(Ruthven & Llamas, 2003). The set-theoretic models such as the Boolean model represent documents as sets of words of phrases whilst algebraic models such as the Vector Space Model represent documents and queries usually as vectors, matrices or tuples. Probabilistic models such as language models, on the other hand, perform probabilistic inference to determine whether a document is relevant for a given query(Dong et al., 2008). The Boolean model is rarely utilized in recent times, even though its simple methodology saves time and cost of computing, possibly because it is based on a binary criterion, which accepts only exact-matches and not fuzzy or semantic matches. The vector space model is more appropriate as it accepts partial matching of query-document terms through the IDF weighting scheme whilst with language models, documents are ranked in decreasing order of their probability of being relevant. Language models are probably preferred as it directly models the idea that a document is a good match to a query if the document model is likely to generate the query(Dong et al., 2008; Manning et al., 2008). Even so, query expansion models embedded in language models have only shown minimal increase in retrieval performance. It is worth examining whether the choice of retrieval model plays a role in effectiveness of the query expansion model. Thus, query expansion models should be assessed not only with language models but also with the Boolean and Vector Space Modes. Each query expansion pattern requires that the query is modified by adding a certain category of terms into the original query. Our opinion is that it is necessary to ensure that each category of terms is assigned weights in order to express their level of significance within the expanded query. We propose that original query terms and their synonyms are the most important as they represent the exact user intent. The next most significant candidate terms for a query would be the semantically related terms, i.e. meronyms, hyponyms, hypernyms and coordinate terms and lastly, the co-occurring terms as they represent directly and indirectly relevant terms, respectively. However, in each process that generates potential expansion terms, the terms are assigned statistical values and ranked according to some measures of importance, referred to local analysis. Depending on the process involved for the extraction of related terms from multiple sources, varying statistical values would be generated, thus requiring a reconciliation scheme that analyzes, on a global level, the associations between the categories of terms pooled.

Measuring the importance of a term is typically assessed statistically within a document corpus (Nanas et al., 2004). Two popular term weighting methods are the inverse-document frequency (IDF) computation that discriminates relevant documents based on the appearance of terms across documents within the document corpus (Jones, 2004) and probabilistic weighting schemes that predict the relevance of a document by estimating the presence or absence of a term within a document (Robertson & Sparck-Jones, 1976). The IDF scheme is widely used whilst the probabilistic relevance prediction schemes have gained prominence in recent years through the use of probabilistic information retrieval models such as BM25 (Robertson & Zaragoza, 2009). It is important to note that the choice of the term weighting scheme for query expansion is very much dependent on the retrieval model utilized

## 8.3 Perspectives: Towards Integrated Query Expansion Models

Having discussed related literature, we observe that both external knowledge-based and statistical processing bring about benefits to the query expansion process. External knowledge based processing allows for directly related terms (semantically related) to be obtained while statistical processing allows for indirectly related terms (co-occurring terms) be extracted to ensure all relevant documents can be retrieved for a given query. Due to the data sparsity problem, researchers have highlighted that even with large amounts of data, significant evidence of potential term relationships are not fully captured. Thus, the following research efforts have attempted to integrate multiple sources of evidence to predict term relations more accurately.

### 8.3.3 Early Attempts

Three research efforts  Cao et al(Cao et al., 2005) & Collins-Thompson & Callan(Collins-Thompson & Callan, 2005) perform integration of statistical processing and external knowledge based processing. In Figure 8, we depict the process flow of the proposed integrated query expansion models. An original query is subjected to fundamental linguistic processing of stemming and stop word removal. The remaining terms serve as input for term dependency modelling in statistical processing and external knowledge based processing approaches for expansion concepts extraction. The extracted concepts are reconciled via term weighting models that are embedded within an existing retrieval model.

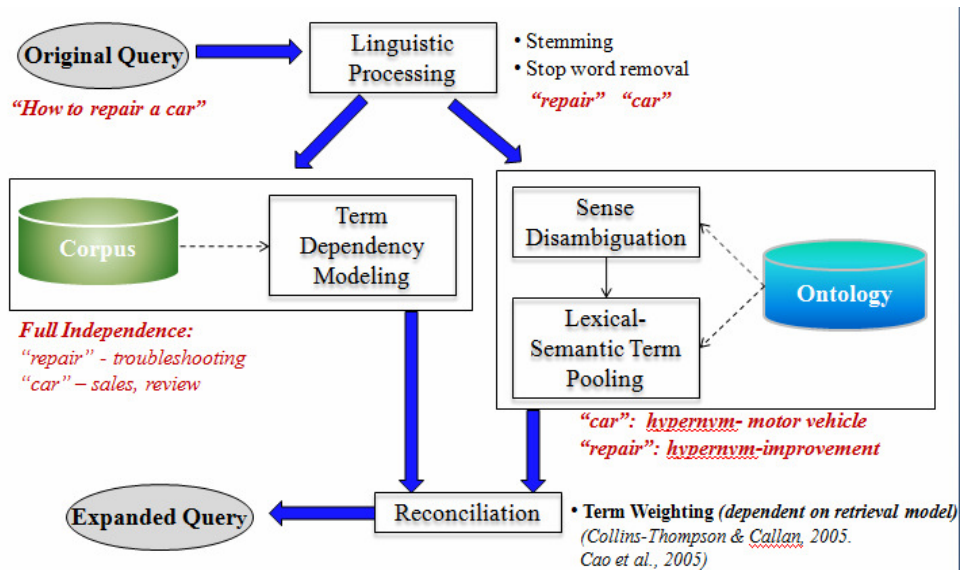

**Figure 8 Integrated Query Expansion Model**

Cao et al. (2005) propose an approach that integrates terms from two sources, the co-occurring terms in document corpus and three types of related terms i.e. synonymy, hyponymy and hypernymy from the WordNet ontology ("WordNet", 2006). The authors claim that the most important terms in information retrieval are nouns and as such choose to focus on three relations of nouns in WordNet . Given that different types of relations should not play the same role, the link model of this system is separated into several sub-models to adequately represent the dissimilarity between the type of relations in the ontology. WordNet does not assign weights to any relations between terms it defines. As such, the authors rely on the text collection by counting the co-occurrence of these words in the collection to estimate the conditional probability between words, with an imposed condition that the words must co-occur within a window of certain size. As many pairs of words in the vocabulary do not have a link in Wordnet, relative frequency of co-occurrences alone is not sufficient for this estimation and as such the authors utilize a smoothing method. The weight estimation model utilized in this approach is similar to the principle of pseudo-relevance feedback which assumes the top *n* documents to be relevant to the query, thus having higher weights. The results of their experiments show that the unigram model, co-occurrence model and link model perform moderately on their own and the most significant increase in precision and recall is seen when the three models are combined. The difference between this approach and previous work is the introduction of the link model that highlights the varying significance of WordNet relations. The results obtained from the experiments show that the relations contained in WordNet can well complement the co-occurrences statistics extracted from the collection. The retrieval model utilized in this framework is the language model.

Collins-Thompson & Callan (2005) propose a Markov chain framework that estimates the likelihood of relevance of potential expansion terms through the combination of multiple sources of knowledge. A chain of inference is built to represent different types of evidence at different random walk stages. The features used by the random walk come from a variety of sources, such as term co-occurrence in an external corpus, co-occurrence in the top retrieved documents, synonym dictionaries, general word association scores, and so on. Potential expansion terms which are closely related to the main aspects of a query are given more preference even if they are not as rare in the collection as other expansion terms. Statistically significant differences in accuracy were observed depending on the weighting of evidence in the random walk. For example, using co-occurrence data later in the walk was generally better than using it early. The word associations are obtained through experiments that required up to 6000 persons to write the first word that came to mind for a word presented to them. The first word that comes to mind is assumed to be meaningfully related or strongly associated to the presented word. This is a potentially useful resource to further gauge the context and relations of terms that could be used for query expansion. The retrieval model utilized in this framework is the language model and the Indri Search Engine is used for indexing and retrieval. The latter, which is part of the Lemur toolkit (Available at http://www.lemurproject.org/indri/) combines a language modelling approach with inference networks. The results show that using co-occurrence from top-ranked retrieved documents along with general word association, synonyms, and stemming shows an increase in retrieval performance.

Xu et al. (2006) conducted a comparative study utilizing three loosely-coupled existing query expansion techniques (e.g. local analysis, global analysis and ontology-based term re-weighting) for Bio-Medical literature search. Global analysis entails the extraction of related term co-occurrences from the entire document collection whilst local analysis involves the extraction of highly-related terms from the preliminary result set obtained through the processing of the initial user query. For local analysis, the top co-occurrence terms for expansion are obtained through the use of the Latent Semantic Indexing (LSI) algorithm and Association Rule (AR). In contrast, in global analysis, terms to be expanded are extracted from all the documents in the collection based on the medical term co-occurrence frequencies provided by the Unified Medical Language System, a domain-specific ontology which provides medical concepts and their related co-occurrence frequencies. Experimental runs were performed using the vector space and probabilistic models on the Lemur Toolkit (2008) with the probabilistic model showing better precision in retrieval results. The authors conclude that assigning different weights for original and extended terms according to their relative importance in queries provides the best results among the three query expansion strategies and adding only one extended term to each original query term of the LSI local analysis increases precision over adding multiple terms.

### 8.3.4 Analysis & Open Issues

Among these approaches, Cao et al. (2005) and Collins-Thompson & Callan (2005) perform integration of statistical processing and external knowledge based processing. Xu et al. (2006), on the other hand, essentially perform statistical based processing and utilizes a domain specific ontology to provide contextual information for term weighting. It does not actually incorporate terms extracted from the ontology for query expansion. It is evident in these approaches that query structures and varying query lengths are not differentiated and linguistic characteristics of a query not scrutinized extensively. As in all other query expansion approaches, queries are stemmed and stop words removed prior to the process of expansion. Collins-Thompson & Callan's approach differs from the others in terms of the sources of expansion terms that they choose to integrate. They indeed consider lexically related terms, i.e. synonyms, morphological variants, word associations and co-occurring terms. They do not integrate however semantically related terms that can be obtained from external knowledge sources. Cao et al., on the other hand, only integrate two sources which are the co-occurring terms and the semantically related terms from the WordNet ontology. Even though these approaches integrate a variety of expansion terms, they do not differentiate the query expansion patterns. An important factor for consideration in the integration of multiple sources is the way in which weights are assigned to the different types of potential expansion terms. Unlike other approaches, Cao et al. estimate weights for each category of expansion terms differentiating their level of importance within a given query. We believe that this differentiation has an impact on the retrieval performance as formally defined related terms (semantically related) would possibly be more significant to an information need than proximity based related terms (co-occurring). Also, Cao et al's approach integrates both types of related terms and has shown the most increase in precision and recall compared to approaches that do not integrate both types of terms or attempt to utilize them independently. However, compared to the baseline unigram model, the increase in the average precision and recall is still marginal and could possibly be improved by incorporating all factors outlined pertaining to query expansion and by improving the way in which the multiple sources are reconciled. Cao et al's and Collins-Thompson & Callan's expansion models are embedded into a language model. . It would be of interest to see whether the use of a different retrieval model will have a better impact on retrieval performance.

## 9. Evaluation & Efficiency of Query Expansion Models

Several evaluation and performance issues are important to note in the proposal and implementation of query expansion models such as maximization of precision/recall and computational efficiency. As described in Section 8.1, maximizing precision requires inclusion of precise terms whilst maximizing recall requires inclusion of more general terms in the expanded query. The effectiveness of query expansion techniques have typically been evaluated in accordance to improvements observed in precision. The integrated query expansion frameworks (Collins-Thompson & Callan, 2005; Cao et .al,2005) report a minimal increase in precision ranging from 4-6% over baseline systems This minimal increase is possibly due to the lack of consideration for linguistic characteristics as well as query expansion patterns were not considered when formulating the expanded query. The latter may also explain the reason for degradation of retrieval performance (Metzler & Croft, 2007) seen in some queries when subjected to query expansion. Thus, our proposal of considering query expansion patterns during the expansion term extraction and selection process is very likely to result in appropriate selection of evaluation measures.

It is important to consider computational efficiency when proposing query expansion frameworks. Real-time application of query expansion models require not only minimal user interaction (described in Section 1) but also minimal turn-around time. Thus, computationally intensive applications should be avoided to ensure easy utilization in the real-world. However, discussion of computational efficiency is non-existent in current query expansion literature.

We suggest a two-pronged approach to evaluation of query expansion models i.e.system-based and user-based. System-based evaluation will allow for preliminary validation of the proposed system as well as assessment of computational efficiency. User-based evaluation would ensure user satisfaction and provide valuable insights for better automation of the query expansion process.

## 10. Conclusion

From a review of related literature, factors that influence the performance of query expansion methods have been identified. To recapitulate the findings, a summary of these factors are presented in this section highlighting the research issues that require further investigation.

Users express their information need through natural language queries which are not always well structured and may be semantically ambiguous. Depending on their familiarity with the search process, users construct queries which are either short and straight to the point or long-winded. These queries may take the form of grammatically correct sentences or merely a group of keywords associated to their search goals. This variance in expressing information need points to the need to study the structure and length of queries and identify the kind of information that can be extracted to understand the information need expressed.

Queries are made up of a series of words that are grouped on the basis of morphological, syntactical and semantic properties. Morphological analysis of query terms would allow for its root form and variants to be derived. Syntactical analysis results in the tagging of the query terms based on its POS and the identification of syntactical relations among them. Syntactical relations between query terms would provide semantic information that leads to the identification of query concepts that represent the query intent. Semantic ambiguities that arise from synonymous and polysemic query terms may be resolved by understanding contextual information from adjacent concepts. Further study of these linguistic characteristics is warranted to decipher the intent of a query accurately.

It is important to recognize that the vocabulary utilized in a document to represent a topic of interest may differ from those expressed in a query. A document can be deemed as relevant if it contains the exact words that are used in a query or if it contains terms which are lexically or semantically related to the original query. There are also terms that may be relevant to a topic if they frequently co-occur with a query term in a document within certain proximity. This set of related terms can be obtained from external knowledge bases and through the statistical processing of a document corpus. Further investigation is needed to improve or to select the most appropriate technique of relevant term extraction from an ontology or document corpus.

As a result of linguistic processing, statistical processing and knowledge based processing, multiple potential expansion terms would be generated. To eliminate redundant terms in the various categories of potential terms, concept mapping strategies would be required to filter terms which are semantically identical or equivalent. It is necessary to propose a fusion framework to integrate potential expansion terms which are obtained from multiple sources. To emulate the sort of expansion that is typically performed by users, query expansion patterns (i.e. enhancement with synonyms, generalization, specialization and parallel movement) should be considered when designing term selection and weighting schemes. Lastly, since there exists multiple retrieval models with different benefits and drawbacks, the impact of retrieval models on the performance of query expansion models is worthy of exploration.

Table 1 illustrates the essential factors that have been incorporated in existing query expansion models. It is evident from this tabulation that none of the existing research efforts fully integrate all the necessary factors, thus emphasizing a novel research direction in the area of query expansion.

|  | Linguistic Analysis | | | External Knowledge Based Processing | Statistical Processing | Reconciliation of Multiple Expansion Term Sources | Distinction of Query Expansion Patterns | Utilization of Multiple Retrieval Models |
|---|---|---|---|---|---|---|---|---|
|  | Morphological | Syntactical | Semantic | | | | | |
| (Voorhees, 1994) | √ |  | √ | √ |  |  |  |  |
| (Liu et al., 2004) | √ |  | √ | √ |  |  |  |  |
| (Gao et al., 2004) | √ |  |  |  | √ |  |  |  |
| (Bai et al., 2005) | √ |  |  |  | √ |  |  |  |
| (Cao et al., 2005) | √ | √ | √ | √ | √ |  |  |  |
| (Collins-Thompson & Callan, 2005) | √ | √ | √ | √ | √ |  |  |  |
| (Radhouani et al., 2006) |  |  | √ | √ |  |  |  |  |
| (Xu et al., 2006) | √ |  |  | √ | √ |  |  | √ |
| (Abdelali et al., 2007) | √ |  |  |  | √ |  |  |  |
| (Metzler & Croft, 2007) | √ |  |  |  | √ |  |  |  |
| (Lioma & Ounis, 2008) |  | √ |  |  | √ |  |  |  |
| (Song et al., 2008) | √ | √ |  |  | √ |  |  |  |

Table 1 Summary of Query Expansion Factors Incorporated in Existing Systems

**References:**

Abdelali, A., Cowie, J., & Soliman, H. S. (2007). Improving query precision using semantic expansion. *Information Processing & Management, 43*(3), 705-716.

Aula, A. (2003). Query Formulation in Web Information Search. *Proceedings of IADIS International Conference WWW/Internet.* , 403-410.

Arampatzis, A., & Kamps, J. (2008). A study of query length. *Proceedings of the 31st Annual International ACM SIGIR Conference on Research and Development in Information Retrieval*, 811-812.

Avi, A., & Jaap, K. (2008). A study of query length. *Proceedings of the 31st Annual International ACM SIGIR Conference on Research and Development in Information Retrieval*, 811-812

Bai, J., Song, D., Bruza, P., Nie, J.-Y., & Cao, G. (2005). Query expansion using term relationships in language models for information retrieval. *Proceedings of the 14th ACM International Conference on Information and Knowledge Management*, 688 - 695.

Baeza-Yates, R. & Ribeiro-Neto, B. (1999), Modern Information Retrieval. Addison-Wesley

Bhogal, J., Macfarlane, A., & Smith, P. (2007). A review of ontology based query expansion. *Information Processing & Management, 43*(4), 866-886.

Bozzon, A., Chirita, P.-A., Firan, C. S., & Nejdl, W. (2007). Lexical analysis for modeling web query reformulation. *Proceedings of the 30th annual international ACM SIGIR conference on Research and development in information retrieval*, 739 - 740.

Broder, A. (2002). A Taxonomy of Web Search. *SIGIR Forum 36*(2), 3-10.

Budanitsky, A., & Hirst, G. (2006). Evaluating WordNet-based Measures of Lexical Semantic Relatedness. *Comput. Linguist., 32*(1), 13-47.

Bendersky, M., Croft, W.B., and Smith, D.A. 2009. Two-stage query segmentation for information retrieval.*ACM SIGIR*, 810-811.

Cao, G., Nie, J.-Y., & Bai, J. (2005). Integrating word relationships into language models. *Proceedings of the 28th annual international ACM SIGIR conference on Research and development in information retrieval*, 298 - 305.

Chirita, P.A., Firan, S.F., & Nejdl, W. (2007). Personalized query expansion for the web. *Proceedings of the 30th annual international ACM SIGIR conference on Research and development in information retrieval*, 7 - 14.

Collins-Thompson, K., & Callan, J. (2005). Query expansion using random walk models. *Proceedings of the 14th ACM Intl conference on Information and knowledge management*, 704 - 711.

Croft, W. B., Turtle, H. R., & Lewis, D. (1991). The use of phrases and structured queries in information retrieval. *Proc. 14th Ann. Intl. ACM SIGIR Conf. on Research and Development in Information Retrieval*, 32-45.

Dong, H., Hussain, F. K., & Chang, E. (2008). A survey in traditional information retrieval models. *2nd IEEE International Conference on Digital Ecosystems and Technologies.*, 397-402.

Efthimis N. Efthimiadis. Query Expansion. *Annual Review of Information Systems and Technology ARIST*, Vol. 31 (1996) , p. 121-187.

Fagan, J. L. (1987). Automatic phrase indexing for document retrieval: An examination of syntactic and non-syntactic methods. *Proceedings of the 10th Annual International ACM SIGIR Conference on Research and Development in Information Retrieval*, 91-101.

Frakes,W.B., Baeza-Yates, R. (1992) Information Retrieval: Data Structures & Algorithms . Prentice Hall.Gao, J., Nie, J.-Y., Wu, G., & Cao, G. (2004). Dependence language model for information retrieval. *Proceedings of the 27th annual international ACM SIGIR conference on Research and development in information retrieval*, 170 - 177.

Goldsmith, J. (2001). Unsupervised learning of the morphology of a natural language. *Comput. Linguist., 27*(2), 153-198.

Google N-gram.. from http://www.ldc.upenn.edu/Catalog/CatalogEntry.jsp?catalogId=LDC2006T13.

Hollink, L., Schreiber, A. T., Wielinga, B. J., & Worring, M. (2004). Classification of user image descriptions. *Int. J. Hum.-Comput. Stud., 61*(5), 601-626.

Huston, S. and Croft, W.B.,2010. Evaluating verbose query processing techniques. *ACM SIGIR*, 291 298.

Cui, H., Wen, J-R., Nie,J-Y, Ma, W-Y.(2003). Query expansion by mining user logs. IEEE Transactions on Knowledge & Data Engineering. 829-839.

Jansen, B. J., Booth, D. L., & Spink, A. (2008). Determining the informational, navigational, and transactional intent of Web queries. *Information Processing & Management*, 1251-1266.

Jose, J., Furner, J., & Harper, D. J. (1998). Spatial querying for image retrieval: a user-oriented evaluation. *Proceedings of the 21st annual international ACM SIGIR conference on Research and Development in Information Retrieval*, 232 - 240.

Jones, K. S. (2004). A statistical interpretation of term specificity and its application in retrieval. Journal of Documentation, 60(5), 493 - 502.

Kim, S.-B., Seo, H.-C., & Rim, H.-C. (2004). Information retrieval using word senses: root sense tagging approach. *Proceedings of the 27th annual international ACM SIGIR conference on Research and development in information retrieval*, 258 - 265.

Lang, H., Wang, B., Jones, G., Li, J., & Xu, Y. (2008). An Evaluation and Analysis of Incorporating Term Dependency for Ad-Hoc Retrieval *30th European Conference on IR Research*, 602-606.

Lau, E. P., & Goh, D. H.-L. (2006). In search of query patterns: A case study of a university OPAC. *Information Processing & Management, 42*(5), 1316-1329.

Lease, M. 2009. An improved markov random field model for supporting verbose queries. *ACM SIGIR,* 476-483.

Lin, D. (1999). Automatic identification of non-compositional phrases. *Proceedings of the 37th annual meeting of the Association for Computational Linguistics on Computational Linguistics*, 317 - 324.

Link Grammar. (2008). from http://www.link.cs.cmu.edu/link/.

Lioma, C., & Ounis, I. (2008). A syntactically-based query reformulation technique for information retrieval. *Information Processing & Management, 44*(1), 143-162.

Liu, S., Liu, F., Yu, C., & Meng, W. (2004). An effective approach to document retrieval via utilizing WordNet and recognizing phrases. *Proceedings of the 27th annual international ACM SIGIR conference on Research and development in information retrieval*, 266 - 272.

Liu, S., Yu, C., & Meng, W. (2005). Word sense disambiguation in queries. *Proceedings of the 14th ACM international conference on Information and knowledge management*, 525 - 532.

Manning, C. D., Raghavan, P., & Schütze, H. (2008). *Introduction to Information Retrieval*: Cambridge University Press.

Marneffe, M.-C. d., & Manning, C. D. (2008). Stanford Typed Dependencies Manual. from http://nlp.stanford.edu/software/dependencies_manual.pdf.

Mastora, A., Monopoli, M., & Kapidakis, S. (2008). Exploring Query Formulation and Reformulation: A Preliminary Study to Map Users' Search Behaviour. *Proceedings of the 12th European conference on Research and Advanced Technology for Digital Libraries*, 427 - 430.

McCarthy, D., & Carroll, J. (2003). Disambiguating Nouns, Verbs, and Adjectives Using Automatically Acquired Selectional Preferences. *Computational Linguistics, 29*(4), 639-654.

McCarthy, D., Koeling, R., Weeds, J., & Carroll, J. (2007). Unsupervised acquisition of predominant word senses. *Computational Linguistics, 33*(4), 553-590.

McDonald, S., & Tait, J. (2003). Search strategies in content-based image retrieval. *Proceedings of the 26th annual international ACM SIGIR conference on Research and development in information retrieval*, 80 - 87.

Mena, E., Kashyap, V., Illarramendi, A., & Sheth, A. P. (2000). Imprecise Answers in Distributed Environments: Estimation of Information Loss for Multi-Ontology Based Query Processing. *International Journal of Cooperative Information Systems, 9*(4), 403-425.

Metzler, D., & Croft, W. B. (2005). A Markov random field model for term dependencies. *Proceedings of the 28th annual international ACM SIGIR conference on Research and development in information retrieval*, 472 - 479.

Metzler, D., & Croft, W. B. (2007). Latent concept expansion using markov random fields. *Proceedings of the 30th annual international ACM SIGIR conference on Research and development in information retrieval*, 311 - 318.

Miller, G. A., Beckwith, R., Fellbaum, C., Gross, D., & Miller, K. J. (1990). Introduction to WordNet: An On-line Lexical Database. *International Journal of Lexicography, 3*(4), 235-244.

Moreau, F., Claveau, V., & Sébillot, P. (2007). *Automatic Morphological Query Expansion Using Analogy-Based Machine Learning* Paper presented at the 29th European Conference on IR Research, Rome, Italy, 222-233

Morenberg, M. (2001). *Doing Grammar* (Vol. Third Edition): Oxford University Press.

Nanas, N., Uren, V., & de Roeck, A. (2004). A comparative evaluation of term weighting methods for information filtering. Proceedings of the 15th International Workshop on Database and Expert Systems Applications, 13-17.

Patwardhan, S., Banerjee,S., and Pedersen,T. (2007) UMND1: Unsupervised Word Sense Disambiguation Using Contextual Semantic Relatedness. *SemEval-2007.* 390-393,

Pedersen, T., Patwardhan, S., & Michelizzi, J. (2004). WordNet::Similarity - Measuring the relatedness of Concepts. *American Association for Artificial Intelligence (AAAI)*.

Peng, F., Ahmed, N., Li, X., & Lu, Y. (2007). Context sensitive stemming for web search. *Proceedings of the 30th annual international ACM SIGIR conference on Research and development in information retrieval*, 639 - 646.

Phan, N., Bailey, P., & Wilkinson, R. (2007). Understanding the relationship of information need specificity to search query length. *Proceedings of the 30th annual international ACM SIGIR conference on Research and development in information retrieval*, 709 - 710.

Punyakanok, V., Roth, D., & Yih, W.-t. (2008). The importance of syntactic parsing and inference in semantic role labeling. *Computational Linguistics, 34*(2), 257-287.

Mihalcea, R (2007). Using Wikipedia for Automatic Word Sense Disambiguation. (NAACL 2007). 196-203.

Radhouani, S., Lim, J. H., Chevallet, J.-P., & Falquet, G. (2006). Combining Textual and Visual Ontologies to Solve Medical Multimodal Queries. *IEEE International Conference on Multimedia and Expo 2006*, 1853 - 1856.

Rieh, S. Y., & Xie, H. (2006). Analysis of multiple query reformulations on the web: The interactive information retrieval context. *Information Processing & Management, 42*(3), 751-768.

Robertson, S. & Sparck-Jones, K. (1976). Relevance weighting of search terms. Journal of the American Society for Information Science, 27(3), 129-146.

Robertson, S. & Zaragoza, H. (2009). The Probabilistic Relevance Framework: BM25 and Beyond. Foundations and Trends in Information Retrieval, 3(4), 333-389.

Rose, D. E., & Levinson, D. (2004). Understanding user goals in web search. *Proceedings of the 13th international conference on World Wide Web*, 13 - 19.

Salton, G., & Buckley, C. (1990). Improving retrieval performance by relevance feedback. *Journal of the American Society for Information Science*, 355-364 .

Song, R., Taylor, M. J., Wen, J.-R., Hon, H.-W., & Yu, Y. (2008). Viewing Term Proximity from a Different Perspective *30th European Conference on IR Research*, 346-357.

Spink, A., Greisdorf, H., & Bateman, J. (1998). From highly relevant to not relevant: examining different regions of relevance. *Information Processing Management, 34*(5), 599-621.

Stanford NLP Parser. from http://nlp.stanford.edu/software/lex-parser.shtml

Stevenson, M., & Wilks, Y. (2001). The interaction of knowledge sources in word sense disambiguation. *Computational Linguistics, 27*(3), 321-349.

TREC. Text REtrieval Conference. from http://trec.nist.gov/overview.html.

Turney, P. D. (2006). Similarity of Semantic Relations. *Comput. Linguist., 32*(3), 379-416.

Vechtomova, O. (2009) Query Expansion for Information Retrieval. Encyclopedia of Database Systems. 2254-2257

Voorhees, E. M. (1994). Query expansion using lexical-semantic relations. *Proceedings of the 17th annual international ACM SIGIR conference on Research and development in information retrieval*, 61-69.

Wang, Y.-C., Vandendorpe, J., & Evens, M. (1985). Relational thesauri in information retrieval. *Journal of the American Society for Information Science, 36*(1), 15-27.

Westman, S., Lustila, A., & Oittinen, P. (2008). Search strategies in multimodal image retrieval. *Proceedings of the 2nd international symposium on Information interaction in context*, 13-20

White, R. W., & Morris, D. (2007). Investigating the querying and browsing behavior of advanced search engine users. *Proceedings of the 30th annual international ACM SIGIR conference on Research and development in information retrieval*, 255 - 262.

Wolfram, D. (2008). Search characteristics in different types of Web-based IR environments: Are they the same? *Information Processing & Management, 44*(3), 1279-1292.

Xu, X., Zhu, W., Zhang, X., Hu, X., & Song, I.-Y. (2006). A Comparison of Local Analysis, Global Analysis and Ontology-based Query Expansion Strategies for Bio-medical Literature Search. *Systems, Man and Cybernetics, 2006. SMC '06. IEEE International Conference on, 4*, 3441-3446.

Yinghao, L., et al., (2007). Improving weak ad-hoc queries using wikipedia as external corpus, ACM SIGIR.